\documentclass[a4paper,10pt]{article}
\usepackage{amsmath,amssymb}

\newtheorem{theorem}{Theorem}[section]

\newtheorem{lemma}[theorem]{Lemma}

\allowdisplaybreaks[1]

\font\msytw=msbm10 scaled\magstep1
\font\msytww=msbm8 scaled\magstep1

\font\indbf=cmbx10 scaled\magstep2

\let\a=\alpha \let\b=\beta    \let\g=\gamma     \let\d=\delta     \let\e=\varepsilon
  \let\h=\eta     \let\th=\vartheta \let\k=\kappa     \let\l=\lambda
\let\m=\mu    \let\n=\nu                      \let\r=\rho
\let\s=\sigma \let\t=\tau            
\let\ps=\psi   \let\o=\omega     
 \let\D=\Delta       \let\L=\Lambda    
             
\let\O=\Omega 

\def\VV{{\cal V}}
\def\WW{{\cal W}}

\def\DD{{\cal D}}

\def\pp{{\bf p}}\def\xx{{\bf x}}
\def\yy{{\bf y}}\def\kk{{\bf k}}\def\nn{{\bf n}}
\def\zz{{\bf z}}

       \def\oo{{\underline \omega}}
\def\ee{{\underline \varepsilon}}

\def\u0{{\underline 0}}

        \def\ZZZ{\hbox{\msytw Z}}
\def\zzzz{\hbox{\msytww Z}}

\let\dpr=\partial

\let\==\equiv

\let\io=\infty
\let\0=\noindent

\def\*{{\hfill\break\null\hfill\break}}

\def\ie{\hbox{\it i.e.\ }}
\def\eg{\hbox{\it e.g.\ }}

\def\tilde#1{{\widetilde #1}}

\def\la{{\langle}}
\def\ra{{\rangle}}

\def\tende#1{\,\vtop{\ialign{##\crcr\rightarrowfill\crcr
             \noalign{\kern-1pt\nointerlineskip}
             \hskip3.pt${\scriptstyle #1}$\hskip3.pt\crcr}}\,}
\def\otto{\,{\kern-1.truept\leftarrow\kern-5.truept\to\kern-1.truept}\,}
\def\fra#1#2{{#1\over#2}}

\def\Tr{\rm Tr}

\def\wh#1{\widehat{#1}}
\def\hat#1{\wh{#1}}
\def\sqt[#1]#2{\root #1\of {#2}}

\def\VV{{\cal V}}
\def\WW{{\cal W}}

\def\DD{{\cal D}}

\def\T#1{{#1_{\kern-3pt\lower7pt\hbox{$\widetilde{}$}}\kern3pt}}
\def\VVV#1{{\underline #1}_{\kern-3pt
\lower7pt\hbox{$\widetilde{}$}}\kern3pt\,}
\def\W#1{#1_{\kern-3pt\lower7.5pt\hbox{$\widetilde{}$}}\kern2pt\,}

\def\indica{\leaders \hbox to 0.5cm{\hss.\hss}\hfill}
\def\guida{\leaders\hbox to 1em{\hss.\hss}\hfill}
\mathchardef\oo= "0521

\def\pp{{\bf p}}\def\xx{{\bf x}}
\def\yy{{\bf y}}\def\kk{{\bf k}}\def\nn{{\bf n}}
\def\zz{{\bf z}}

\def\oo{{\underline \omega}}

\def\qed{\raise1pt\hbox{\vrule height5pt width5pt depth0pt}}

\def\indic{\hbox{\raise-2pt \hbox{\indbf 1}}}

 \def\ZZZ{\hbox{\msytw Z}}
\def\zzzz{\hbox{\msytww Z}}

\def\lb#1{\label{#1}}

\def\be{\begin{equation}}
\def\ee{\end{equation}}
\def\bea{\begin{eqnarray}}\def\eea{\end{eqnarray}}
\def\bean{\begin{eqnarray*}}\def\eean{\end{eqnarray*}}
\def\bfr{\begin{flushright}}\def\efr{\end{flushright}}
\def\bc{\begin{center}}\def\ec{\end{center}}
\def\bal{\begin{align}}
\def\ba#1{\begin{array}{#1}}
\def\bd{\begin{description}}\def\ed{\end{description}}
\def\bv{\begin{verbatim}}\def\ev{\end{verbatim}}
\def\bsp{\begin{split}}
\def\nn{\nonumber}
\def\Halmos{\hfill\vrule height10pt width4pt depth2pt \par\hbox to \hsize{}}
\def\pref#1{(\ref{#1})}
 
\def\virg{\,,\quad}

\begin{document}

\title{Drude weight in non solvable quantum spin chains}

\author{
G. Benfatto\thanks{Dipartimento di Matematica, Universit\`a di Roma ``Tor
Vergata'', Via della Ricerca Scientifica, I-00133, Roma}
\\e-mail: benfatto@mat.uniroma2.it,
\and V. Mastropietro{${}^\ast$}
\\e-mail: mastropi@mat.uniroma2.it
}

\date{December 1, 2010}
\maketitle

\begin{abstract}
For a quantum spin chain or 1D fermionic system, we prove that the {\em Drude
weight} $D$ verifies the universal Luttinger liquid relation $v_s^2=D/\k$,
where $\k$ is the {\em susceptibility} and $v_s$ is the {\em Fermi velocity}.
This result is proved by rigorous Renormalization Group methods and is true for
any weakly interacting system, regardless its integrablity. This paper,
combined with \cite{BM1}, completes the proof of the Luttinger liquid
conjecture for such systems.
\end{abstract}

\section{Introduction and main result}

Quantum spin chains and one dimensional Fermi systems have been the subject of
an intense theoretical investigation for decades, either for their remarkable
properties or for the fact that they can be experimentally realized in systems
like quantum spin chain models (KCuF3) \cite{L} or carbon nanotubes \cite{Au}.
We consider a quantum spin chain model with Hamiltonian
\be\lb{1.1} H=-\sum_{x=1}^{L-1} [S^1_x S^1_{x+1}+S^2_x
S^2_{x+1}] -h \sum_{x=1}^L S^3_x +\l \sum_{1\le x,y\le L} v(x-y) S^3_x S^3_{y}+
U^1_L\;, \ee
where $S^\a_x = \s^\a_x/2$ for $i=1,2,\ldots,L$ and $\a=1,2,3$, $\s^\a_x$ being
the Pauli matrices, and $U^1_L$ is a boundary term; finally $v(x)=v(-x)$ and
$|v(x)|\le C e^{-\k |x|}$.

If $v(x-y)=\d_{|x-y|,1}$ and $h=0$, \pref{1.1} is the hamiltonian of the $XXZ$
spin chain in a zero magnetic field, which can be diagonalized by the Bethe
ansatz \cite{YY}. No exact solution is known for more general interactions, but
some particular models have been the subject of an extensive numerical
analysis.

It is well known that the quantum spin model can be equivalently written in
terms of fermionic anticommuting operators $a_x^\pm \= \prod_{y=1}^{x-1}
(-\s_y^3) \s_x^\pm$; if $J_1=J_2=1$, one gets
\be\lb{z}
\bsp H = &-\sum_{x=1}^{L-1} \fra12 [ a^+_{x} a^-_{x+1}+ a_{x+1}^+
a^-_{x}] -h\sum_{x=1}^L ( a^+_x a^-_x-{1\over 2}) \\
&+\l \sum_{1\le x,y\le L} v(x-y)( a^+_x a^-_x-{1\over 2})( a^+_{y}
a^-_{y}-{1\over 2}) +U_L^2\virg \end{split}
\ee
where $U_L^2$ is the boundary term in the new variables. We choose it so that
the fermionic Hamiltonian coincides with the Hamiltonian of a fermion system on
the lattice with periodic boundary conditions.

%If $\h$ is the exponent of the 2-point function (see (4) of
%\cite{BM1}), $\bar\n$ is the correlation length exponent (see (11)
%of \cite{BM1}), $X_+$ is the density exponent (see (7-9) of
%\cite{BM1}) and $X_-$ is the Copper pair exponent (see (10) of
%\cite{BM1}), some of the relations conjectured in \cite{K,LP}for
%the model \pref{1.1} are the following
%
%\be X_+=K\quad\quad X_-=K^{-1}\quad\quad\bar\n={1\over
%1-K^{-1}}\quad\quad 2\h=K+K^{-1}-2\label{1a} \ee
%
%A number of arguments has been given for justifying the validity
%of \pref{1a} has been proposed along the years \cite{PB,N,ZZ,S},
%but only in recent times a rigorous proof of them has been reached
%\cite{M,BFM,BM1}, based on the hidden local symmetries and Ward
%Identities; in particular, it was proved in Theorem 1 of
%\cite{BM1} that, for $\l$ small enough, there exists an {\it
%analytic} function $K(\l)$ such that \pref{1a} are true, with
%
%\be K(\l)=1-\l {\hat v(0)-\hat v(2p_F)\over \pi\sin
%p_F}+O(\l^2)\label{2a}\ee with $\cos p_F=h+\l$. Note that the
%exact relations \pref{1a} are {\it universal} \ie do not depend
%from the Hamiltonian details, for instance from the form of the
%interaction $v(\vec x)$, contrary to the function $K(\l)$
%\pref{2a}.
%In addition to universal relations between the critical exponents
%like \pref{1a}, Haldane proposed the validity of relations
%connecting the Fermi velocity $v_s$ (see (4) of \cite{BM1}), the
%susceptibility $\k$ and the {\it Drude weight} $D$ which are
%defined in the following way.

If $O_x$ is a local monomial in the $S^\a_x$ or $a^\pm_x$ operators, we call
$O_\xx=e^{H x_0} O_x e^{-H x_0}$ where $\xx=(x_0,x)$ and $x_0$ is the
``imaginary time''; moreover, if $A=O_{\xx_1}\cdots O_{\xx_n}$
\be <A>_{L,\b}= {\Tr[e^{-\b H}{\bf T}(A)]\over \Tr[e^{-\b H}]}\virg\ee
${\bf T}$ being the time order product, denotes its expectation in the grand
canonical ensemble, while $<A>_{T;L,\b}$ denotes the corresponding truncated
expectation. We will use also the notation $<A>_T=\lim_{L,\b\to\io}
<A>_{T;L,\b}$.

The {\it response functions} measure the response of the system to an external
probe. In particular, the spin conductivity properties of model \pref{1.1} can
be obtained in the model \pref{z} from the {\it current-current} response
function, whose Fourier transform is defined as
\be \hat G^{0,2}_{J,J}(\pp)=
\lim_{\b\to\io}\lim_{L\to\io}\int_{-\b/2}^{\b/2} dx_0\sum_{x\in\L} e^{i\pp\xx}
\la J_\xx J_{\bf 0}\ra_{T;L,\b}\label{yy}\virg \ee
where $\pp=(p_0,p)$, $p_0={2\pi\over\b}n, p={2\pi\over L}m$, $(n,m)\in \ZZZ^2$,
$-[L/2]\le m\le [(L-1)/2]$, $J_\xx=e^{H t} J_x e^{-H t}$ and  $J_x$ is the
paramagnetic part of the current
\be J_x={1\over 2i} [a_{x+1}^+ a^-_{x} - a^+_{x} a^-_{x+1}]\,. \ee
A crucial quantity in the study of the conductivity properties is played by the
{\it Drude weight}, defined in the following way. Let us consider the function
\be \hat D(\pp)=-\D -\hat G^{0,2}_{J,J}(\pp)\virg\label{88} \ee
where $\D= <\D_x>$ and
\be\D_x= - {1\over 2} [a^+_{x} a^-_{x+1}+a_{x+1}^+ a^-_{x}] \ee
is the diamagnetic part of the current, whose mean value $<\D_x>$ is indeed
independent of $x$, hence it is equal to $<H_T>/L$, with $H_T= \sum_x \D_x$,
the value of $H$ for $h=\l=0$. Then the Drude weight is given by
\be D=\lim_{p_0\to 0} \hat D(p_0,0)\,.\label{89} \ee
If one assumes analytic continuation in $p_0$ around $p_0=0$, one can compute
the conductivity in the linear response approximation by the Kubo formula, see
\eg \cite{Ma}, that is
\be
\s = \lim_{\o\to 0}\lim_{\d\to 0}{\hat D(-i \o+\d,0)\over -i \o+\d}\,.
\ee
Therefore, a nonvanishing $D$ indicates infinite conductivity.

Another important quantity is the {\em susceptibility}, which can be
calculated, in the fermionic representation, in terms of the {\it
density-density} response function $G^{0,2}_{\r,\r}(\xx)=\la \r_\xx \r_{\bf
0}\ra_T$, $\r_x=a^+_x a_x$, by the equation
\be\lb{89a}\k=\lim_{p\to 0}\hat G^{0,2}_{\r,\r}(0,p)\virg \ee
where $\hat G^{0,2}_{\r,\r}(\pp)$ is defined analogously to \pref{yy}. Note
that, in the fermionic representation, $\k= \k_c \r^2$, where $\k_c$ is the
{\it fermionic compressibility} and $\r$ is the fermionic density, see \eg
(2.83) of \cite{[8a]}.

The large distance behavior of the response functions is given (for coupling
not too large) by power laws with non-universal exponents depending on all
details of the Hamiltonian, like the form of the potential and the value of the
magnetic field. Only for the interaction $v(x-y)=\d_{|x-y|,1}$ a solution is
known by Bethe ansatz \cite{YY}, if $h=0$; by using this explicit solution,$\k$
and $D$ can be computed. However, even in that case, only a single exponent can
be calculated \cite{Ba}.

In \cite{BM,BMa,BMb} rigorous RG methods have been applied to spin chains or
fermi\-onic 1D systems, regardless their integrability; the outcome of such
analysis is that several physical observables, and in particular the critical
exponents, can be written as convergent series. The exponents are
interaction-dependent but nevertheless verify universal model independent
relations; if $\h$ is the exponent of the 2-point function (see \eg (4) of
\cite{BM1}), $\bar\n$ is the correlation length exponent (see \eg (11) of
\cite{BM1}), $X_+$ is the density exponent (see \eg (7-9) of \cite{BM1}) and
$X_-$ is the Copper pair exponent (see \eg (10) of \cite{BM1}), it has been
proved in \cite{BFM,BM1} that, for $\l$ small enough,
\be X_+=K\virg X_-=K^{-1}\virg \bar\n={1\over 1-K^{-1}}\virg
2\h=K+K^{-1}-2\virg\label{1a} \ee
where $K(\l)$ is an analytic function such that
\be K(\l)=1-\l {\hat v(0)-\hat v(2p_F)\over \pi\sin
p_F}+O(\l^2)\virg\label{2a}\ee
with $\cos p_F=-h-\l$.
Note that the exact relations \pref{1a} are {\it universal}, \ie do not depend
on the Hamiltonian details, for instance on the form of the interaction $v(x)$,
contrary to the function $K(\l)$, see \pref{2a}.

Universal relations connect also the critical exponents with the susceptibility
$\k$; in \cite{BM1} it was proved that
\be \k={K\over \pi v_s}\,.\label{ggg1} \ee
In this paper we prove the following Theorem for the Drude weight.

\begin{theorem}\lb{th1}
If $\l$ is small enough, the function $\hat D(\pp)$ defined in \pref{88} can be
written, for $\pp$ small but different from $0$, in the form
\be \hat D(\pp)= {v_s\over \pi}K {p_0^2\over p_0^2+v_s^2 p^2}
+H(\pp)\virg\ee
where $H(\pp)$ is a continuous function, such that $|H(\pp)|\le C|\pp|^{\th}$,
with $0<\th<1$; therefore the Drude weight is given by
\be D={v_s K\over \pi} \label{ggg}\ee
and satisfies the identity
\be\lb{Llr} v^2_s=D/\k\,. \ee
\end{theorem}
The validity of the relations \pref{1a}, \pref{ggg1},\pref{ggg} and \pref{Llr}
is the content of the {\it Luttinger liquid conjecture} formulated in \cite{Ha}
(see also \cite{K,LP}); given the Drude weight and the susceptibility, one can
determine exactly all the exponents and the Fermi velocity. All these relations
are true in the {\it Luttinger model}, describing interacting fermions with a
relativistic linear dispersion relation and solved by bosonization \cite{ML};
the content of the Luttinger liquid conjecture is that they are true also in
the model \pref{1.1}, even if the exponents are completely different. This is
by no means obvious; the exponents, $\k$ and $D$ are non universal functions of
the interaction, and surely depend on the dispersion relation and the details
of the Hamiltonian. The validity of the conjecture was partially checked on the
solvable $XXZ$ chain; $v_s$, $\k$, $D$ can be computed from the Bethe ansatz
solution \cite{Ha,GS} and the validity of the relation \pref{Llr} (following
from \pref{ggg},\pref{ggg1}) is verified. Moreover, by using \pref{1a}, the
exponents can be exactly determined from the knowledge of $\k$ and $D$; note
that the value of $\bar\n$ found in this way agrees with the one obtained in
\cite{Ba}. A number of arguments have been proposed along the years
\cite{PB,N,S} in order to justify the validity of \pref{1a}, \pref{ggg1} and
\pref{ggg}, but they rely on unproved assumptions or approximations in non
solvable cases.

The present paper completes the proof of the Luttinger liquid conjecture for
quantum spin chain or 1D fermionic system with generic weak short range
interaction. The proof relies on a number of technical results previously
established in \cite{BM,BMa,BMb,BFM,BM1}; in particular the present paper
extends and completes the analysis of \cite{BM1}, which we assume the reader
familiar with.

The Drude weight in quantum spin chains has been the subject in recent years of
an intense numerical investigation \cite{Z,AG,KEH,RA,FK,HS}, with the main
objective of detecting a possible different behavior of conductivity at finite
temperatures between the integrable and the non integrable cases; it has been
conjectured that the Drude weight is non vanishing also at finite temperature
in the integrable cases, while it is vanishing in non integrable systems, but
the results are still controversial. Our methods for the calculation of the
Drude weight at zero temperature can be applied either to solvable or non
solvable systems, and we believe that an extension of these methods would allow
us to understand also the properties of the Drude weight at non zero
temperature.

\section{Ward Identities}

We shall proceed as in App. B of \cite{BM1}. Let us consider the (imaginary
time) conservation equation:
\be\lb{eqm}
{\partial \r_\xx\over \partial x_0}= e^{H x_0} [H,\r_x] e^{-H x_0}
=-i\dpr^{(1)}_x J_{\xx} \= -i [J_{x,x_0}-J_{x-1,x_0}]\;,
\ee
where we have used that $[H,\r_x]=[H_T,\r_x]$, $H_T$ being the value of $H$ for
$h=\l=0$. This equation implies some exact identities involving various
correlation functions, that play the role in the lattice models of the usual
Ward Identities (WI) of continuous relativistic models. They are valid at any
finite $\b$ and $L$, but we shall use them only in the limit $L=\b=\io$.

We shall call $G^{2}(\xx,\yy) = <a^-_\xx a^+_\yy>$ the (imaginary time) Green's
function, while
$$G^{2,1}_\r(\xx,\yy,\zz) = <\r_\xx a^-_\yy
a^+_\zz>_T \mbox{\ and\ \ } G^{2,1}_J(\xx,\yy,\zz) = <J_\xx a^-_\yy
a^+_\zz>_T$$
will be the vertex functions. By using \pref{eqm} one gets the WI
\be
\bsp &{\dpr\over \dpr x_0} G^{2,1}_\r(\xx,\yy,\zz)  = -i \dpr^{(1)}_x
G^{2,1}_J(\xx,\yy,\zz)+\\
+\d(x_0 &- z_0) \d_{x,z} G^2(\yy,\xx) - \d(x_0-y_0)\d_{x,y}
G^2(\xx,\zz)\label{ref1}\;, \end{split}
\ee
where $\dpr^{(1)}_x$ is the lattice derivative. In the same way a WI for the
density-density correlations is derived. If we define
$$G^{0,2}_{\r,\r}(\xx,\yy) = <\r_\xx \r_\yy>_T,\;
G^{0,2}_{\r,J}(\xx,\yy) = <\r_\xx J_\yy>_T,\;
G^{0,2}_{J,J}(\xx,\yy) = <J_\xx J_\yy>_T,$$
we get
\be\bsp
{\dpr\over \dpr x_0} G^{0,2}_{\r,\r}(\xx,\yy)  &= - \dpr^{(1)}_x
G^{0,2}_{J,\r}(\xx,\yy) + \d(x_0-y_0) <[\r_{(x,x_0)}\virg
\r_{(y,x_0)}]>,\\
{\dpr\over \dpr x_0} G^{0,2}_{\r,J}(\xx,\yy)  &= -i \dpr^{(1)}_x
G^{0,2}_{J,J}(\xx,\yy) + \d(x_0-y_0) <[\r_{(x,x_0)}, J_{(y,x_0)}]>\,.
\end{split}
\ee
Noting that $[\r_{(x,x_0)}, \r_{(y,x_0)}]=0$, while
\be
[\r_{(x,x_0)}, J_{(y,x_0)}]= -i\d_{x,y}\D_{(x,x_0)} +i\d_{x-1,y}\D_{(y,x_0)},
\ee
we get, using that $<\D_\xx>=<\D_x>=\D$,
\be\lb{1000}
\bsp -i p_0  \hat G^{0,2}_{\r,\r}(\pp) &-i(1-e^{-ip})
\hat G^{0,2}_{J,\r}(\pp) =0\virg\\
-i p_0  \hat G^{0,2}_{\r,J}(\pp) &-i(1-e^{-ip}) \hat G^{0,2}_{J,J}(\pp)
=i(1-e^{-ip}) \D\;. \end{split}
\ee
Hence, by using the definition \pref{88}, the WI \pref{1000} and the fact that
$\hat G^{0,2}_{\r,J}(\pp) = \hat G^{0,2}_{J,\r}(-\pp)$, we get
\be p_0^2\, \hat G^{0,2}_{\r,\r}(\pp) -4 \sin^2 (p/2\,) \hat D(\pp)=0\;. \ee
The above equation holds quite generally for fermionic lattice systems. If
$\hat G^{0,2}_{\r,\r}(\pp)$ and $\hat D(\pp)$ were continuous in $\pp=0$, it
would imply that both $\k$ and $D$ are vanishing. In the case we are
considering, we will see in the next section that $\hat G^{0,2}_{\r,\r}(\pp)$
and $\hat D(\pp)$ are bounded but not continuous in $\pp=0$, which is
sufficient to prove only that:
\be \hat G^{0,2}_{\r,\r}(p_0,0)=0\virg \hat
D(0,p)=0\label{ref}\;. \ee

\section{Renormalization Group anaysis}
It is well known that the correlations of the quantum spin chain can be derived
by the following Grassmann integral, see \S 2.1 of \cite{BM}:
\be\lb{1z} e^{\WW_{L,\b,M}(A,J,\phi)}=\int P(d\psi) e^{-\VV(\psi)+
B(A,J,\psi) + \int d\xx [\phi^+_\xx \psi^-_{\xx} + \psi^-_{\xx} \psi^+_\xx]}\;,
\ee
where $\cos p_F=-\l-h-\n$, $v_s=v_F(1+\d)$, $\psi^\pm_\xx$ and $\phi^\pm_\xx$
are Grassmann variables, $\int d\xx$ is a shortcut for
$\sum_{x}\int_{-\b/2}^{\b/2} dx_0$, $P(d\psi)$ is a Grassmann Gaussian measure
in the field variables $\psi^\pm_\xx$ with covariance (the free propagator)
given by
\be
g_M(\xx-\yy)= {1\over\b L}\sum_{\kk\in\DD_{L,\b}}  {\chi(\g^{-M} k_0) e^{i\d_M
k_0} e^{i\kk(\xx-\yy)} \over -i k_0+ (v_s/v_F) (\cos p_F -\cos k)}\;,
\ee
where $\chi(t)$ is a smooth compact support function equal to $0$ if $|t|\ge
\g>1$ and equal to $1$ for $|t|<1$, $\kk=(k,k_0)$, $\kk\cdot\xx=k_0x_0+kx$,
${\cal D}_{L,\b}\={\cal D}_L \times {\cal D}_\b$, ${\cal D}_L\=\{k={2\pi n/L},
n\in \zzzz, -[L/2]\le n \le [(L-1)/2]\}$, ${\cal D}_\b\=\{k_0=2(n+1/2)\pi/\b,
n\in Z\}$ and
\be\bsp
\VV(\psi) &=\l \int d\xx d\yy \tilde v(\xx-\yy) \psi_\xx^+ \psi_\yy^+
\psi_\yy^- \psi_\xx^-+ \n \int d\xx \psi_{\xx}^+\psi_{\xx}^- \;-\\
&- \d \int d\xx [\cos p_F \psi_{\xx}^+ \psi_{\xx}^- -(\psi_{\xx+{\bf
e_1}}^+\psi_{\xx}^- + \psi_\xx^+ \psi_{\xx+ {\bf e_1}}^-)/2]\virg \end{split}
\nn\ee
with ${\bf e_1}=(0,1)$, $\tilde v(\xx-\yy)=\d(x_0-y_0)v(x-y)$. Moreover
\be\label{1.9}
\bsp B(A,J,\psi) = \int &d\xx \Big\{\psi^+_{\xx}\psi^-_\xx A_0(\xx) + {1\over
2i} [\psi^+_{\xx+{\bf e}_1} \psi^-_{\xx} - \psi^+_{\xx} \psi^-_{\xx+{\bf e}_1}]
A_1(\xx) -\\
&-{J\over 2}[\psi^+_{\xx} \psi^-_{\xx+{\bf e}_1}+
\psi_{\xx+{\bf\e}_1}^+\psi^-_{\xx}]\big)\Big\}\,.
\end{split}
\ee
Note that, due to the presence of the ultraviolet cut-off $\g^M$, the Grassmann
integral has a finite number of degree of freedom, hence it is well defined.
The constant $\d_M=\b/\sqrt{M}$ is introduced in order to take correctly into
account the discontinuity of the free propagator $g(\xx)$ at $\xx=0$, where it
has to be defined as $\lim_{x_0\to 0^-} g(0,x_0)$; in fact our definition
guarantees that $\lim_{M\to\io} g_M(\xx)=g(\xx)$ for $\xx\not=0$, while
$\lim_{M\to\io} g_M(0,0)=g(0,0^-)$. The density and current correlations can be
written in terms of functional derivatives of \pref{1z}
\bal G^{0,2}_{\r,\r}(\xx,\yy) &= \lim_{\b\to\io} \lim_{L\to\io} \lim_{M\to\io}
\frac{\d^2}{\d A_0(\xx) \d A_0(\yy)} W_{L,\b,M}(A,0,0)\big|_{A=0}\virg\nn\\
G^{0,2}_{\r,J}(\xx,\yy) &= \lim_{\b\to\io} \lim_{L\to\io} \lim_{M\to\io}
\frac{\d^2}{\d A_0(\xx) \d A_1(\yy)} W_{L,\b,M}(A,0,0)\big|_{A=0}\virg\nn\\
\\[-28pt]
\nn\\
G^{0,2}_{J,J}(\xx,\yy) &= \lim_{\b\to\io} \lim_{L\to\io} \lim_{M\to\io}
\frac{\d^2}{\d A_1(\xx) \d A_1(\yy)} W_{L,\b,M}(A,0,0)\big|_{A=0}\virg\nn\\
\D &= \lim_{\b\to\io}\lim_{L\to\io}\lim_{M\to\io}{1\over \b L}\frac{\d}{\d J}
W_{L,\b,M}(0,J,0)\big|_{J=0}\;.\nn
\end{align}
In \cite{BM,BMa,BMb} a multiscale integration procedure combined with Ward
Identities allows us to write the above correlations in terms of a convergent
expansion; the counterterms $\n,\d$ are chosen so that $p_F$ is the Fermi
momentum and $v_s$ is the Fermi velocity. By using Theorem 3.12 of \cite{BM},
one can easily prove that $\D$ is a finite constant. However, the bounds
obtained from the multiscale analysis for $G^{0,2}_{\r,\r}(\xx,\yy)$ and
$G^{0,2}_{J,J}(\xx,\yy)$ are not sufficient to prove that their Fourier
transforms are bounded around $\pp=(0,0)$. In fact, by using theorem (1.5) of
\cite{BM}, we see that their non-oscillating part behaves for large $|\xx-\yy|$
as $|\xx-\yy|^{-2}$, so that logarithmic divergences in the Fourier transform
cannot be excluded.

In order to compute the Fourier transform of the current-current correlation we
will follow the same strategy used in \cite{BM1} for the density-density
correlation. We introduce a continuous model with linear dispersion relation
regularized by a non local fixed interaction, together with ultraviolet $\g^N$
and an infrared $\g^l$ momentum cut-offs. The model is expressed in terms of
the following Grassmann integral:
\be\lb{vv1}
\bsp
e^{\WW_{N}(J,\tilde J,\phi)} &= \int\! P_Z(d\psi) e^{-\VV^{(N)}(\sqrt{Z}\psi +
\sum_{\o=\pm} \int\! d\xx [Z^{(3)}
J_{\xx} +\o\, \tilde Z^{(3)} \tilde J_{\xx}] \r_{\xx,\o}}\cdot\\
&\cdot e^{Z \sum_{\o=\pm} \int d\xx [\psi^{+}_{\xx,\o} \phi^-_{\xx,\o} +
\phi^+_{\xx,\o} \psi]}\virg
\end{split}
\ee
where $\r_{\xx,\o} = \psi^{+}_{\xx,\o} \ps^{-}_{\xx,\o}$, $\xx\in\tilde\L$ and
$\tilde\L$ is a square lattice of side $L$, whose size is of order $\g^{-l}$,
say $\g^{-l}/2 \le L \le \g^{-l}$; $P_Z(d\psi^{[l, N]})$ is the fermionic
measure with propagator
\be\lb{gth} {1\over Z} g_{th,\o}(\xx-\yy)={1\over Z}{1\over
L^2}\sum_{\kk}e^{i\kk\xx}{\chi_{N}(\kk)\over -ik_0+\o c k}\;, \ee
where $Z$ and $c$ are two parameters, to be fixed later, and $\chi_{l,N}(\kk)$
is a cut-off function depending on a small positive parameter $\e$,
nonvanishing for all $\kk$ and reducing, as $\e\to 0$, to a compact support
function equal to $1$ for $\g^{l}\le |\kk|\le \g^{N+1}$ and vanishing for
$|\kk|\le \g^{l-1}$ or $|\kk|\ge \g^{N+1}$ (its precise definition can be found
in (21) of \cite{BMa}); moreover, the interaction is
\be\lb{gjhfk} \VV^{(N)}(\psi)={\l_\io\over 2} \sum_{\o}\int d\xx
\int d\yy v_0(\xx-\yy) \psi^+_{\xx,\o}
\psi^-_{\xx,\o}\psi^+_{\yy,-\o}\psi^-_{\yy,-\o}\;, \ee
where $v_0(\xx-\yy)$ is a rotational invariant potential, of the form
\be v_0(\xx-\yy)={1\over L^2}\sum_{\pp} \hat v_0(\pp)
e^{i\pp(\xx-\yy)}\;, \ee
with $|\hat v_0(\pp)|\le C e^{-\m |\pp|}$, for some constants $C$, $\m$, and
$\hat v_0(0)=1$. We define

\bal
&G^{2,1}_{th,\r;\o}(\xx,\yy,\zz) =\lim_{-l,N\to\io}\lim_{a^{-1},L\to\io}
{\partial\over\partial J_\xx} {\partial^2\over
\partial\phi^+_{\yy,\o} \partial\phi^-_{\zz,\o}}
\WW_{l,N}(J,\tilde J,\phi)|_{J=\tilde J=\phi=0}\virg\nn\\
&G^{2,1}_{th,J;\o}(\xx,\yy,\zz) =\lim_{-l,N\to\io}\lim_{a^{-1},L\to\io}
{\partial\over\partial \tilde J_\xx} {\partial^2\over
\partial\phi^+_{\yy,\o} \partial\phi^-_{\zz,\o}}
\WW_{l,N}(J,\tilde J,\phi)|_{J=\tilde J=\phi=0}\virg\nn\\
&G^{2}_{th;\o}(\yy,\zz) = \lim_{-l,N\to\io}\lim_{a^{-1},L\to\io}
{\partial^2\over \partial\phi^+_{\yy,\o} \partial\phi^-_{\zz,\o}}
\WW_{l,N}(J,\tilde J,\phi)|_{J=\tilde J=\phi=0}\virg\\
&G^{0,2}_{th,\r,\r}(\xx,\yy) = \lim_{-l,N\to\io} \lim_{a^{-1},L\to\io}
{\partial^2\over \dpr J_\xx \dpr J_\yy}
\WW_{l,N}(J,\tilde J,\phi)|_{J=\tilde J=\phi=0}\virg\nn\\
&G^{0,2}_{th,J,J}(\xx,\yy)
=\lim_{-l,N\to\io}\lim_{a^{-1},L\to\io}{\partial^2\over \dpr \tilde J_\xx \dpr
\tilde J_\yy} \WW_{l,N}(J,\tilde J,\phi)|_{J=\tilde J=\phi=0}\,.\nn
\end{align}
The existence of the $N\to\io$ limit has been proved in \cite{M} and in \S 3 of
\cite{BFM}, extending the method used in \cite{Le} for the analysis of the
Yukawa model in two dimensions; the existence of the limit $l\to-\io$ has been
proved in \cite{BM,BMa,BMb}.

The model \pref{vv1} is a sort of {\it effective model} for the lattice
fermionic model (2); it is indeed well known that a non relativistic gas of
fermions in one dimension admits an effective description in terms of massless
Dirac fermions in $d=1+1$ dimension. We can make precise this idea via the
following lemma, whose proof is an immediate extension of the proof given in \S
3 of \cite{BM1} for the density-density correlation.

\begin{lemma}\lb{lm1} Given $\l$ small enough, there exist constants $Z$, $Z^{(3)}$,
$\tilde Z^{(3)}$, $\l_\io$, depending analytically on $\l$, such that
$Z=1+O(\l^2)$, $Z^{(3)}=1+O(\l)$, $\tilde Z^{(3)}=v_F+O(\l)$,
$\l_\io=\l+O(\l^2)$ and, if $c=v_s$ and $|\pp|\le \k \le 1$,
\be\lb{h10}
\bsp \hat G^{0,2}_{\r,\r}(\pp) &= \hat G^{0,2}_{th,\r,\r}(\pp)+
A_{\r,\r}(\pp)\;,\\
\hat G^{0,2}_{J,J}(\pp) &= \hat G^{0,2}_{th,J,J}(\pp) + A_{J,J}(\pp)+ \D\virg
\end{split}
\ee
with $A_{\r,\r}(\pp),A_{J,J}(\pp)$ Lipschitz continuous in $\pp$. Moreover, if
we put $\pp_F^\o=(0,\o p_F)$ and we suppose that $0<\k\le
|\pp|,|\kk'|,|\kk'-\pp|\le 2\k$, $0<\th<1$, then
\be\lb{h10a}
\bsp \hat G^{2,1}_\r(\kk'+ \pp_F^\o, \kk'+\pp+\pp_F^\o)
&= \hat G^{2,1}_{th,\r;\o}(\kk',\kk'+\pp)[1+O(\k^\th)]\;,\\
\hat G^{2,1}_J(\kk'+ \pp_F^\o, \kk'+\pp+\pp_F^\o)
&= \hat G^{2,1}_{th,J;\o}(\kk',\kk'+\pp)[1+O(\k^\th)]\;,\\
\hat G^2(\kk'+\pp_F^\o) &= \hat G^{2}_{th,\o}(\kk')[1+O(\k^\th)]\;.
\end{split}
\ee
\end{lemma}

This lemma says that the vertex functions of the two models are essentially
coinciding close to the Fermi momenta, if the bare parameters are chosen
properly, while the response functions differ by a continuous function. Note
also the the bare parameters of the model \pref{vv1} are expressed by
convergent expansions depending on all model details, but the WI imply that
they are not independent parameters, as we will see shortly.

The main reason behind the introduction of the model \pref{vv1} is that, while
the model \pref{1.1} is invariant only under the phase transformation
$\psi^{\pm}_\xx\to e^{\pm i\a} \psi^{\pm}_\xx$, the model \pref{vv1} is invariant under
{\it two} phase transformations, the total $\psi^{\pm}_{\xx,\o}\to e^{\pm i\a}
\psi^{\pm}_{\xx,\o}$ and the chiral $\psi^{\pm}_{\xx,\o}\to e^{\pm \o i\a}
\psi^{\pm}_{\xx,\o}$. This implies that the Fourier transforms of the response
functions can be completely determined from the WI, see app. A of \cite{BM1};
if $D_\o(\pp)=-i p_0+\o c p$, we get:
\be\lb{ggv1}
\bsp
\hat G^{0,2}_{th,J,J} &= {-1\over 4\pi c Z^2}{(\tilde Z^{(3)})^2\over 1-\t^2}
\left[{D_-(\pp)\over D_+(\pp)}+{D_+(\pp)\over D_-(\pp)}+2 \t \right] +O(\pp)\virg\\
\hat G^{0,2}_{th,\r,\r} &= {-1\over 4\pi c Z^2}{( Z^{(3)})^2\over 1-\t^2}
\left[{D_-(\pp)\over D_+(\pp)}+{D_+(\pp)\over D_-(\pp)}-2 \t\right]
+O(\pp)\virg
\end{split}
\ee
where $\t={\l_\io\over 4\pi c}$. Therefore, from \pref{ggv1} and \pref{h10},
since $c=v_s$,
\be
\bsp
\hspace{-.25cm}\hat G^{0,2}_{\r,\r}(\pp) = {-1\over 4\pi v_s Z^2}{(
Z^{(3)})^2\over 1-\t^2} \left[{D_-(\pp)\over D_+(\pp)}+{D_+(\pp)\over
D_-(\pp)}+2 \t\right]
&+A_{\r,\r}(0)+R_\r(\pp),\\
\hspace{-.3cm}\hat D(\pp)= {-1\over 4\pi v_s Z^2}{(\tilde Z^{(3)})^2\over
1-\t^2} \left[{D_-(\pp)\over D_+(\pp)}+{D_+(\pp)\over D_-(\pp)}-2 \t \right]
&+A_{J,J}(0)+ \D
\\[-5pt] &+R_J(\pp)\virg
\end{split}\ee
with $|R_\r(\pp)|,|R_J(\pp)|\le C|\pp|^\th$, $0<\th<1$. The constants
$A_{\r,\r}(0)$, $A_{J,J}(0)$ and $\D$ are expressed by convergent expansions,
but their values can be determined from the WI for the model \pref{1.1};
indeed, by \pref{ggv1}, $\hat G^{0,2}_{th,J,J}$ and $\hat G^{0,2}_{th,\r,\r}$
are not continuous in $\pp=0$, but they are bounded, so that \pref{ref} holds;
this condition fixes the values of $A_{J,J}(0)+\D$ and $A_{\r,\r}(0)$ so that
\be\bsp
\hat G^{0,2}_{\r,\r}(\pp) &= {1\over \pi v_s Z^2}{Z^{(3)})^2\over 1-\t^2} {
v_s^2 p^2\over p_0^2+v_s^2 p^2}+R_\r(\pp)\virg\\
\widehat D(\pp) &= {1\over \pi v_s Z^2}{(\tilde Z^{(3)})^2\over 1-\t^2}
{p_0^2\over p_0^2+v_s^2 p^2}+R_J(\pp)\,.
\end{split}\ee
Moreover, the vertex functions verify the following WI, see (35) of \cite{BM1}:
\be\lb{h11}\bsp
-i p_0 &{Z\over Z^{(3)}} \hat G^{2,1}_{th,\r;\o}(\kk,\kk+\pp)+ \o p\ v_s
{Z\over \tilde Z^{(3)}} \hat G^{2,1}_{th,J;\o}(\kk,\kk+\pp)=\\
&= {1\over 1-\t}[\hat G^{2}_{th;\o}(\kk) - \hat G^{2}_{th;\o}(\kk+\pp)]\,;
\end{split}\ee
hence, by using \pref{h10a} and by comparing \pref{h11} with the WI
\pref{ref1}, we get that the bare parameters are not independent, but verify
the relations:
\be\lb{hh} {Z^{(3)}\over (1-\t)Z}=1\virg v_s\, {Z^{(3)}\over \tilde
Z^{(3)}}=1\virg\ee
implying that
\be\bsp
\hat \O_{\r\r}(\pp) &= {K\over \pi v_s}{ v_s^2p^2\over p_0^2+v_s^2
p^2}+R_\r(\pp)\virg\\
\hat D(\pp) &= {v_s\over \pi}K {p_0^2\over p_0^2+v_s^2 p^2} +R_J(\pp)\virg
\end{split}\ee
with $K={1-\t\over 1+\t}$. Eq. (52) of \cite{BM} shows that $K$ is indeed the
critical index $X_+$, see \pref{1a}; hence, by using \pref{89} and \pref{89a},
we get the relations \pref{ggg} and \pref{ggg1}, which immediately imply
\pref{Llr}, so that Theorem \ref{th1} is proved.


\begin{thebibliography}{999999}

\bibitem{BM1} G. Benfatto, V. Mastropietro, {\it J. Stat. Phys.} {\bf 138},
    1084--1108 (2010).

\bibitem{L} B. Lake {\it et al.}, Nature materials {\bf 4}, 329--334 (2005).

\bibitem{Au} O.M. Auslaender {\it et al.}, Phys. Rev. Lett. {\bf 84},
    1764--1767 (2000); M. Bockrath {\it et al.}, Nature {\bf 397}, 598--601 (1999);
    H. Ishiii {\it et al.}, Nature {\bf 426}, 540--544 (2003);
    %Z. Yao {\it et al},
    %Nature {\bf 401}, 273 (1999).

\bibitem {YY} C.N. Yang, C.P. Yang, {\it Phys. Rev.} {\bf 150}, 321--339
    (1966).

\bibitem{Ma} G.D. Mahan, "{\it Many-Particle Physics}", Kluwer Academic/Plenum,
    New York, 2000.

\bibitem{[8a]} D. Pines, P. Nozieres, {\it The theory of quantum liquids}, W.
    Benjiamin, New York, 1966.

\bibitem{Ba} R. J. Baxter, "{\it Exact Solved Models in Statistical
    Mechanics}", Academic Press, London, 1982.

\bibitem{BM} G. Benfatto, V. Mastropietro, {\it Rev. Math. Phys.} {\bf 13},
    1323--1435 (2001).

\bibitem{BMa} G. Benfatto, V. Mastropietro, {\it Comm. Math. Phys.} {\bf 231},
    97--134 (2002).

\bibitem{BMb} G. Benfatto, V. Mastropietro, {\it Comm. Math. Phys.} {\bf 258},
    609--655 (2005).

\bibitem{BFM} G. Benfatto, P. Falco, V. Mastropietro, {\it Comm. Math. Phys.}
    {\bf 292}, 569--605 (2009); {\it Phys. Rev. Lett.} {\bf 104}, 075701 (2010).

\bibitem {Ha} F.D.M. Haldane, {\it Phys.Rev.Lett.} {\bf 45}, 1358--1362
    (1980); J. Phys. C. {\bf 14}, 2575--2609 (1981).

\bibitem{K} L.P. Kadanoff, {\it Phys. Rev. Lett.} {\bf 39}, 903--906 (1977);
    L.P. Kadanoff, A.C. Brown, {\it Ann. Phys.} {\bf 121}, 318--342 (1979); L.P. Kadanoff, F.
    Wegner, {\it Phys. Rev. B} {\bf 4}, 3989--3993 (1971).

\bibitem{LP} A. Luther, I. Peschel, {\it Phys. Rev. B} {\bf 12}, 3908--3917
    (1975).

\bibitem {ML} D. Mattis, E. Lieb. {\it J. Math. Phys.} {\bf 6}, 304--312
    (1965).

\bibitem{GS} G. G\'omez-Santos, {\it Phys. Rev. B} {\bf 46}, 14217--14218
    (1992).

\bibitem{PB} A.M.M. Pruisken, A.C. Brown, {\it Phys. Rev. B} {\bf 23},
    1459--1468 (1981); A.M.M. Pruisken, L.P. Kadanoff, {\it Phys. Rev. B} {\bf 22}
    5154--5170 (1980).

\bibitem{N} M.P.M. den Nijs, {\it Phys. Rev. B} {\bf 23}, 6111--6125 (1981).

\bibitem{S} H. Spohn, {\it Phys. Rev. E} {\bf 60}, 6411--6420 (1999).

\bibitem{Z} X. Zotos, P. Prelov\v{s}ek {\it Phys. Rev. B} {\bf 53},
    983--986(1996); X. Zotos, {\it Phys. Rev. Lett.} {\bf 82}, 1764--1768 (1998).

\bibitem{AG} J.V. Alvarez, C. Gros, {\it Phys. Rev. B} {\bf 66}, 094403 (2002).

\bibitem{KEH} S. Kirkner, H. Evertz, W. Hanke, {\it Phys. Rev. B} {\bf 59},
    1825--1833 (1998).

\bibitem{RA} A. Rosch, N. Andrei, {\it Phys. Rev. Lett.} {\bf 85}, 1092--1095
    (2000).

\bibitem{FK} S. Fujimoto, N. Kawakami, {\it Phys. Rev Lett.} {\bf 90}, 197202
    (2002).

\bibitem{HS} D. Heidarian, S. Sorella. {\it Phys. Rev. B} {\bf 75}, 241104(R)
    (2007).

\bibitem{M} V. Mastropietro, {\it J. Math. Phys.} {\bf 48}, 022302 (2007).

\bibitem{Le} A. Lesniewski.{\it Comm. Math. Phys.} {\bf 108}, 437--467, (1987).

%\bibitem{[11a]} K. Sano J Phys Soc Jpn, 69, 1000-1003 (2000)

%\bibitem{ZZ} A.B. Zamolodchikov,  Al.B. Zamolodchikov, Soviet Scientific
%    Reviews A {\bf 10}, 269 (1989).

\end{thebibliography}
\end{document}